\documentclass[floatfix,twocolumn,aps,amsmath,amsfonts,showpacs,showkeys]{revtex4}
\usepackage{graphicx,natbib}

\newcommand{\be}{\begin{equation}}
\newcommand{\ee}{\end{equation}}

\newcommand{\ie}{\emph{i.e.\ }}
\newcommand{\cf}{cf.\ }
\newcommand{\kon}{k_{\text{on}}}

\begin{document}

\title{Realistic protein--protein association rates from a
    simple diffusional model neglecting long-range interactions, free
    energy barriers, and landscape ruggedness}

\author{Maximilian Schlosshauer}

\email{MAXL@u.washington.edu}

\affiliation{Department of Physics, University of Washington, Seattle,
  WA 98195}

\author{David Baker} 

\email{dabaker@u.washington.edu} 

\affiliation{Department of Biochemistry, University of Washington,
  Seattle, WA 98195}



\begin{abstract} 
  We develop a simple but rigorous model of protein--protein
  association kinetics based on diffusional association on free energy
  landscapes obtained by sampling configurations within and
  surrounding the native complex binding funnels. Guided by results
  obtained on exactly solvable model problems, we transform the
  problem of diffusion in a potential into free diffusion in the
  presence of an absorbing zone spanning the entrance to the binding
  funnel. The free diffusion problem is solved using a recently
  derived analytic expression for the rate of association of
  asymmetrically oriented molecules. Despite the required high steric
  specificity and the absence of long-range attractive interactions,
  the computed rates are typically on the order of $10^4$--$10^6$
  M$^{-1}$~s$^{-1}$, several orders of magnitude higher than rates
  obtained using a purely probabilistic model in which the association
  rate for free diffusion of uniformly reactive molecules is
  multiplied by the probability of a correct alignment of the two
  partners in a random collision. As the association rates of many
  protein--protein complexes are also in the
  $10^5$--$10^6$~M$^{-1}$~s$^{-1}$, our results suggest that free
  energy barriers arising from desolvation and/or side-chain freezing
  during complex formation or increased ruggedness within the binding
  funnel, which are completely neglected in our simple diffusional
  model, do not contribute significantly to the dynamics of
  protein--protein association. The transparent physical
  interpretation of our approach that computes association rates
  directly from the size and geometry of protein--protein binding
  funnels makes it a useful complement to Brownian dynamics
  simulations.
\end{abstract}

\maketitle

\section{Introduction}

The calculation of rates of protein--protein association is of great
interest to biology. These rates span a wide range of values, from
approximately 10$^3$ to 10$^{10}$~M$^{-1}$~s$^{-1}$. If the two
proteins are modeled as uniformly reactive spheres, the
diffusion-limited rate constant is simply given by the classical
Smoluchowski expression (Smoluchowski 1917), $\kon = 4\pi D R$ (where
D is the relative translational diffusion constant and $R$ is the sum
of the radii), which yields rates of $10^9$--$10^{10}$
M$^{-1}$~s$^{-1}$ for associations relevant to proteins. Usually,
however, proteins exhibit a highly anisotropic distribution of
reactivity over their surface. This can be modelled by localized
reactive sites on the surface of the proteins that have to be
sufficiently precisely aligned for the complex formation to occur.
 
Purely probabilistic models have tried to account for such steric
constraints by multiplying the Smoluchowski rate for uniform spheres
by the probability that, in a random encounter, the two molecules are
properly aligned (``geometric rate'') (Janin (1997) gives an example
of this method). This yields rate constants which are typically
several orders of magnitude lower than the Smoluchowski
diffusion-limited rate and are usually much smaller than the values
experimentally observed for biological complexes. It has been found
that this discrepancy can be moderated by taking into account the
effect of rotational diffusion (Shoup et~al.\ 1981, Northrup and
Erickson 1992); additional rate enhancements are brought about by the
presence of attractive interparticle forces (``electrostatic
steering''; see Schreiber and Fersht 1996, Gabdoulline and Wade 1997,
Vijayakumar et~al.\ 1998), and the formation of a weakly specific,
loosely bound encounter complex that subsequently evolves into the
final bound state (Selzer and Schreiber 1999, Camacho et~al.\ 2000).

To replace the estimation of protein--protein association rates via
the geometric rate by a more accurate method, most authors have
pursued a computational approach by carrying out explicit numerical
simulations of the diffusional association of macromolecules, commonly
referred to as Brownian dynamics (BD) simulations (for an excellent
review, see Gabdoulline and Wade 2002). Here, the protein molecules
are modeled in varying detail, from a simple spherical approximation
up to full atomic detail. In the simulation, the molecules are
initially placed in random orientations at a fixed initial separation
$b$.  Diffusional trajectories, with or without the presence of an
interparticle force (such as electrostatic interactions), are then
generated by means of the Ermak--McCammon algorithm (Ermak and
McCammon 1978). A trajectory is ended either when the molecules have
come together in proper orientation to successfully form a complex, or
when their separation has exceeded a certain truncation value $c>b$
such that the probably for an encounter has become vanishingly small.
The fraction of `successful' trajectories is then used to compute the
association rate $\kon$.

Northrup and Erickson (1992) have used such a BD simulation to
compute the association rate of spherical molecules with a reactive
patch, consisting of four contact points in a $17 \text{\AA} \times 17
\text{\AA}$ square arrangement on a plane tangential to the surface of
the molecules. Reaction is then assumed to occur if three of the four
contact points are correctly matched and within a specified maximum
distance. In the absence of any interparticle forces, the authors find
an association rate of $\kon=10^5$ M$^{-1}$ s$^{-1}$, about two orders
of magnitude higher than the geometric rate. Gabdoulline and Wade
(2001) compute association rates for five protein--protein
complexes using full-atom structures in the presence of long-range
electrostatic forces. The reaction condition is defined by formation
of subsets of the polar contacts observed in the native complex
structure.

\begin{figure}
\includegraphics[scale=0.5]{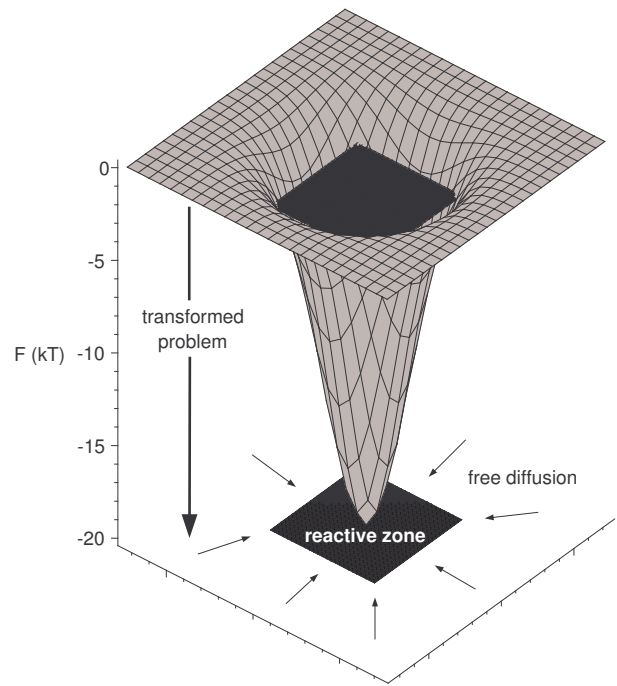}
\caption{\label{fig:funnel}
Simple model of binding dynamics.  Attractive
short-range forces produce a funnel in the free energy landscape
leading into the native complex.  Once the molecules descend several
$kT$ into the funnel, they are effectively captured and binding occurs
rapidly.  In our simple model, the rate of association is approximated
by the rate of free diffusion into a reactive zone in phase space, as
indicated schematically in the $XY$ plane of the drawing.  To compute
the rate of association, we need to first determine the dimensions of
the reactive zone, and second, compute the rate of free diffusion into
this zone.  A more general model would include long-range
(electrostatic) interactions which would bias the diffusion process
towards the funnel entrance.}
\end{figure}

In this paper, we present a different route towards estimation of
rates of bimolecular association. Instead of employing a computer
simulation based approach such as the method of BD simulations
outlined above, we use a recently derived analytical expression
(Schlosshauer and Baker 2002) for the association rate of two
spherical molecules with anisotropic reactivity in the absence of any
interaction forces. The reaction condition is formulated by specifying
the ranges of mutual orientations of the two molecules for which
complex formation will occur. We thus do not require an exact mutual
alignment of the binding partners, but instead assume that favorable
short-range interactions ``guide'' the molecules into their final
bound configurations once the molecules are oriented within specified
angular tolerances (see Fig.~\ref{fig:funnel}). These tolerances can
therefore be viewed as an implicit modelling of attractive short-range
forces. We derive estimates for the tolerances from free energy
landscapes obtained by sampling configurations within and surrounding
the native binding funnel. These values are then used in our
analytical expression to compute the corresponding association rates.
By determining the size and geometry of the aperture in phase space
which must be entered for binding to occur, and rigorously solving the
problem of diffusion through this aperture, our approach provides a
physically transparent complement to BD simulations for computing
binding rates from structures of protein--protein complexes.

\section{Results} 

To compute protein--protein association rates from the
three-dimensional structures of protein--protein complexes according to
the simple diffusive model described above and in Fig.~1, three
ingredients are required. The first is a general theory for computing
the diffusion-limited association rate as a function of the
orientational constraints associated with properly aligigning the two
binding sites (the size and shape of the reactive zone in Fig.~1).
The second is a method for transforming a diffusion in a potential
problem into a free diffusion problem---in the context of Fig.~1, an
estimate of how deeply the reactive zone lies within the binding
funnel (\ie how far molecules must descend into the binding funnel before
they are effectively captured).  The third is a method for mapping the
binding funnel for two proteins given the crystal structure of the
protein--protein complex. We address these issues in Secs.~I, II,
and III, and in Sec.~IV use the results to compute approximate
diffusion limited association rates from the structures of
protein-protein complexes.

\subsection{Theory for the diffusion-limited association rate with
  general orientational constraints}  

Here, we shall restrict ourselves to a brief outline; the full
derivation of our expression for the association rate constant in the
presence of general orientational constraints can be found in
Schlosshauer and Baker (2002).

We consider translational and rotational diffusional motion of two
spherical molecules $A$ and $B$ with radii $R_A$ and $R_B$. To derive
an expression for the association rate constant, we solve the
steady-state translational-rotational diffusion equation describing
the diffusional motion of the two spheres, subject to a reaction
condition that ensures that binding can only occur if the mutual
orientation of the two spheres is sufficiently close to the
orientation in the bound configuration that defines the optimal
alignment.

\begin{figure}
\includegraphics[scale=0.65]{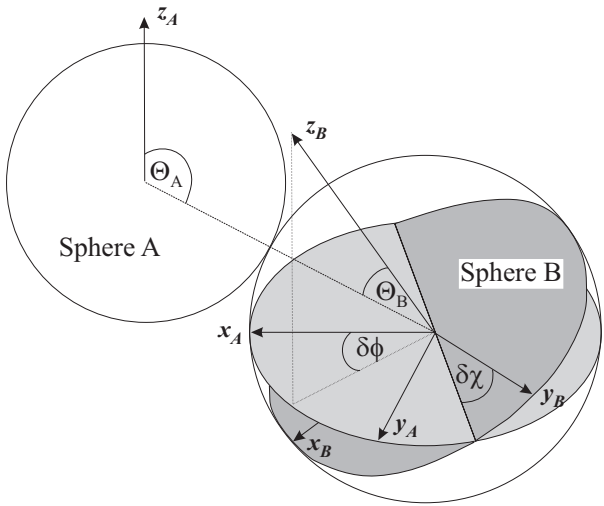}
\caption{\label{fig:model} 
  The axes and angles relevant to the reaction condition,
  Eqs.~\eqref{rc}. The angles $\theta_A$ and $\theta_B$ measure how
  close the center of each reactive patch (coninciding with the
  respective body-fixed $z$ axis) is to the center-to-center vector
  (dashed line).  The angles $\delta \phi$ and $\delta \chi$ denote
  relative torsion angles of the two body-fixed coordinate systems
  $(x_A, y_A, z_A)$ and $(x_B,y_B,z_B)$. For the sake of easier
  visualization of these two angles, the origin of the $x_A$ and $y_A$
  axes (belonging to the coordinate system of sphere $A$) has been
  shifted such as to coincide with the origin of the coordinate system
  of sphere $B$.  Our reaction condition, Eqs.~\eqref{rc}, requires
  near-optimal alignment, \ie all angles $\theta_A$, $\theta_B$,
  $\delta \phi$, and $\delta \chi$ must be below given limits.}
\end{figure}

The reaction conditionis implemented as follows (see
Fig.~\ref{fig:model}): The centers of ``reactive patches'' on the two
spheres are defined by the intersection of the center-to-center vector
with the surfaces of the spheres in the native bound configuration.
Each sphere carries its own body-fixed coordinate system
$\{x_s,y_s,z_s\}$, $s=A,B$, where the $z_s$ axis points at the center
of the reactive patch. The angles $\theta_A$ and $\theta_B$ then
quantify the distance of the center of each reactive patch to the
center-to-center vector, whereas the angles $\delta \phi$ and $\delta
\chi$ denote relative torsional angles between the body-fixed
coordinate systems $(x_A, y_A, z_A)$ and $(x_B,y_B,z_B)$.

At a first glance, one might assume that a fifth parameter is required to
fully describe the mutual orientation of the two spheres---namely,
an azimuthal angle $\phi_A$ in addition to the polar angle $\theta_A$ to
fix the location of the reactive patch on the surface of sphere
$A$. For the formulation of the reaction condition, however, four
angles suffice, because the position
of the center of the reactive patch is automatically specified through
its coincidence with the $z_A$ axis. This leaves 
only one free parameter, namely the ``width'' of the patch,
which is described by the angle $\theta_A$.

The optimal alignment is then defined by $\theta_A=\theta_B=\delta
\phi = \delta \chi = 0$ (additionally, the length $r$ of the center-to-center
vector must be equal to the sum of the radii of the spheres). Our reaction
condition requires that all these angles are suffiently
close to zero for the reaction to occur, \ie that the following
conditions are fulfilled:
\be\label{rc}
\begin{cases}r = R_A + R_B \equiv
R\\
\theta_{A,B} \le \theta_{A,B}^0\\
\delta\phi \le \delta\phi_0\\
\delta\chi \le \delta\chi_0
\end{cases}
\ee
Using the constant-flux approximation introduced by Shoup et~al.\ 
(1991), we obtain for the association rate constant (Schlosshauer and
Baker 2002):
\begin{eqnarray} \label{kon}
\kon &=& D \biggl(\frac{Ra_0}{8\pi^2}\biggr)^2 \Biggl[
\frac{D}{\kappa}a_0 - \nonumber \\ &-& R \sum_{ll_Al_B} 
\frac{K_{l+1/2}(\xi^*)}{l K_{l+1/2}(\xi^*) - \xi^* K_{l+3/2}(\xi^*)} 
\nonumber \\ &\times& q_{ll_Al_B} 
\sum_{n=-l_A}^{+l_A} \biggl( \sum_{m=-l_A}^{+l_A} 
\widehat{C}_{l_Al_B}^{mn} \, \bigl( \begin{smallmatrix} l & l_A & l_B \\ 0 & m & -m 
\end{smallmatrix} \bigr) \biggr)^2 \, \Biggr]^{-1},
\end{eqnarray}
where $D=D_A^{\text{trans}}+D_B^{\text{trans}}$ is the (relative)
translational diffusion constant, $a_0 = (4\pi)^3 \delta\phi_0 \delta\chi_0
(1-\cos\theta_A^0) (1-\cos\theta_B^0)$, $q_{ll_Al_B}=(2l+1)(2l_A+1)(2l_B+1)/16\pi^3$, 
and $\kappa$ quantifies the extent of diffusion control in the
reaction. Furthermore,
\begin{eqnarray}
\widehat{C}_{l_Al_B}^{mn} &=&
\frac{4\pi\sin(m\delta\phi_0)}{m}\frac{4\pi\sin(n\delta\chi_0)}{n}
\\ && \times \, \int_0^{\theta_A^0} \sin\theta_A d\theta_A\, d^{l_A}_{mn}(\theta_A)
\nonumber \\ && \times \, \int_0^{\theta_B^0} \sin\theta_B d\theta_B\,
d^{l_B}_{-m-n}(\theta_B), \nonumber 
\end{eqnarray}
where $d^l_{mn}(\theta)$ denotes the Wigner rotation function. $\bigl(
\begin{smallmatrix} l&l_A&l_B\\m&m_A&m_B 
\end{smallmatrix} \bigr)$ is the Wigner 3-$j$ symbol, and $\xi^* =
R[(D_A^{\text{rot}}/D) l_A(l_A+1) + (D_B^{\text{rot}} / D)
l_B(l_B+1)]^{1/2}$, where $D_A^{\text{rot}}$ and $D_B^{\text{rot}}$
are the rotational diffusion constants. $\widehat{a}_0=a_0/(4\pi\times
8\pi^2 \times 8\pi^2)$ represents the fraction of angular
orientational space over which the reaction can occur, and the
geometric rate is thus given by $\kon = 4\pi D R \times\widehat{a}_0$.

\subsection{Transformation of the diffusion in a potential problem
  into a free diffusion problem} 

A crucial point in the application of Eq.~\eqref{kon} is the
estimation of the angular constraints $\theta_A^0$, $\theta_B^0$,
$\delta\phi_0$, and $\delta\chi_0$. We would like to estimate the
ranges in mutual orientation of the two proteins for which short-range
attractive forces between the atoms are sufficiently dominant to guide
the two molecules into the final bound configuration, and then
translate the problem of diffusional association in the attractive
potential into free diffusion with an absorbing region in
configurational space. To motivate this mapping, we shall first study
two simple toy models for translational and rotational diffusion,
respectively. We then use these ideas to explicitly obtain the angular
constraints for real protein--protein complexes.

\paragraph{Toy model for translational diffusion.}  The reaction rate for
diffusion-controlled bimolecular association of uniformly reactive
spheres in the presence of a potential $U(r)$ can be calculated
exactly and is given by the expression
\be \label{k_pot}
\kon^{(1)} = 4\pi D \biggl[ \int_{R_1}^{\infty} dr\, e^{\beta U(r)}/r^2
\biggr]^{-1}
\ee
where $\beta \equiv \frac{1}{kT}$, and $R_1$ is interpreted as the
center-to-center distance between the associating partners at which
the reaction is assumed to occur. Eq.~\eqref{k_pot} is the classical
result derived by Debye (1942). In the absence of any potential ($U
\equiv 0$), this simplifies to the Smulochowski rate constant for free
diffusion with an absorbing region of width $R_0$, given by
$\kon^{(0)}=4\pi D R_0$ (in the following we use the label ``0'' to
refer to the free diffusion problem, and the label ``1'' to refer to
the diffusion in a potential problem).

It is clear that for $R_0=R_1$, $\kon^{(1)} > \kon^{(0)}$ since the
presence of the (attractive) potential will increase the association
rate. To find the free diffusion analogue of diffusional association
in an attractive potential, we increase $R_0$ until $\kon^{(0)} =
\kon^{(1)}$. In other words, for a given potential, we can determine
the size of the absorbing region (the ``capture radius'') required in
the case of free diffusion to obtain an association rate equivalent to
that of diffusion in the potential. A similar redefinition of the
effective absorbing radius to account for the presence of the
potential was first introduced by Debye (1942).

\begin{figure}
\includegraphics[angle=-90,scale=0.345]{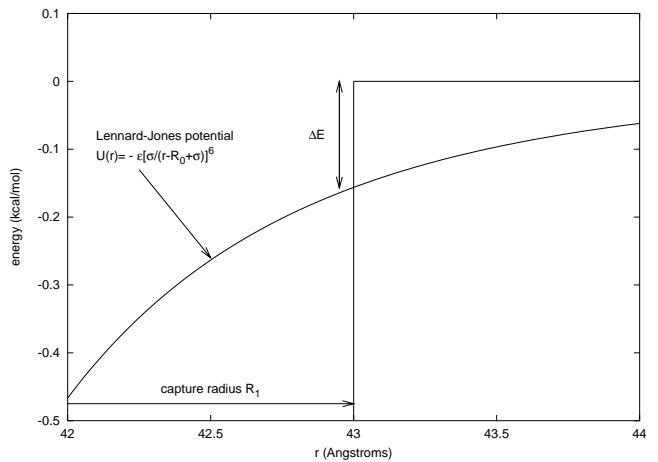}
\caption{\label{fig:pot}
  Mapping of the problem of diffusion in a potential onto that of free
  diffusion with an absorbing region for rotational diffusion on a
  spherical surface.  We use an attractive Gaussian potential
  $U(\theta) = - \epsilon\exp[-(\sigma \theta)^2]$ with $\epsilon =
  10$~kcal/mol and $\sigma = \pi$, where the latter corresponds to a
  (half) width of the potential of
  $\sqrt{\ln(2)}/\sigma=15^{\text{o}}$, a reasonable assumption for a
  short-range potential.  Equating the resulting association rate,
  Eq.~\eqref{eq:theta_pot}, with the rate for free diffusion in
  presence of an absorbing region at $\theta = \theta_0$,
  Eq.~\eqref{eq:theta_free}, we obtain $\theta_0 \approx
  33^{\text{o}}$ for the width of the absorbing region, corresponding
  to an energy drop of $\Delta E \equiv U(\pi)-U(\theta_0) \approx
  0.4$~kcal/mol.}
\end{figure}

A concrete illustration of this procedure is described in
Fig.~\ref{fig:pot} for a Lennard--Jones potential. For a broad range
of parameter values, we find that it is sufficient to drop down by an
energy amount of only $\mathcal{O}(\Delta E) = kT$ to enter the
capture zone, corresponding to less than 5\% of the total depth of the
potential. The capture radius $R_0$ is found to be relatively
insensitive to the depth $\epsilon$ of the potential well, whereas its
dependence on the width $\sigma$ is much stronger---as it must be,
since $R_0$ is an indirect measure of the range of the potential.

The model calculations show that the effect of an attractive potential
$U(r)$ on the association rate can be effectively represented by an
increase in the radius of the interacting spheres, but that in the
relevant case of protein--protein interactions the relative increase
is very small (about 7\% in our example). Since the association rate,
Eq.~\eqref{kon}, is largely insensitive to small changes in the value
of $R$, we conclude that the approximation of using a fixed value
$r=R_A+R_B \equiv R$ for the center-to-center distance of the two
proteins required for the reaction to occur (see our reaction
condition, Eqs.~\eqref{rc}), rather than employing a range of allowed
values (such as demanding that $r \le R_A+R_B + \delta R$ in
Eqs.~\eqref{rc}), is justified.

\paragraph{Toy model for rotational diffusion.} As another important
illustration of our mapping procedure we shall consider
two-dimensional rotational diffusion on a spherical surface in an
attractive Gaussian potential $U(\theta) = - \epsilon
\exp[-(\sigma\theta)^2]$, with $\epsilon > 0$. Again, we would like to
translate this problem into that of free rotational diffusion with an
absorbing region at $\theta = \theta_0$.

Solving the rotational diffusion equation in presence of a potential
$U(\theta)$ yields in the diffusion-controlled limit
\be \label{eq:theta_pot}
\kon^{(1)} = 2\pi D_{\text{rot}} e^{\beta U(\pi)}
\biggl[ \int_{\theta_1}^\pi  d\theta \, 
  e^{\beta U(\theta)} \biggr]^{-1}, 
\ee 
where we let $\theta_1 \rightarrow 0$ for $U \not\equiv 0$ (diffusion
in potential). In the case of free diffusion ($U \equiv 0$),
Eq.~\eqref{eq:theta_pot} becomes
\be \label{eq:theta_free}
\kon^{(0)} = \frac{2\pi D_{\text{rot}} }
{\pi - \theta_0}, 
\ee 
where we now choose $\theta_0 > 0$.  As before, we equate the
association rates, Eqs.~\eqref{eq:theta_pot} and
\eqref{eq:theta_free}, to obtain an estimate for the width $\theta_0$
of the absorbing region.

\begin{figure}
\includegraphics[angle=-90,scale=0.345]{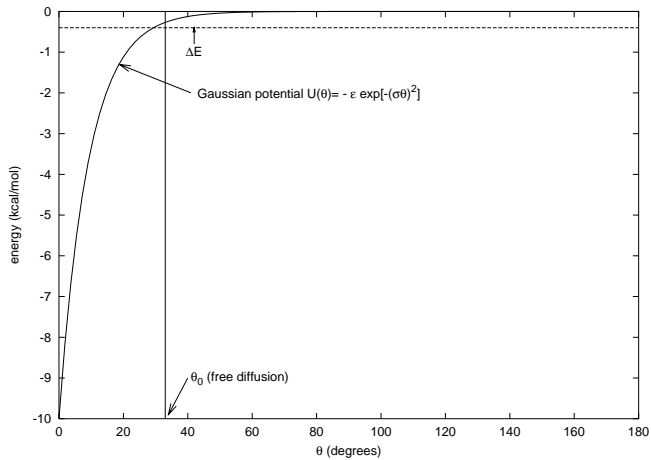}
\caption{\label{fig:rotdiff}Mapping of the problem of 
diffusion in a potential onto that of free diffusion with an absorbing
region for rotational diffusion on a spherical surface.  We use an
attractive Gaussian potential $U(\theta) = - \epsilon\exp[-(\sigma
\theta)^2]$ with $\epsilon = 10$~kcal/mol and $\sigma = \pi$, where
the latter corresponds to a (half) width of the potential of
$\sqrt{\ln(2)}/\sigma=15^{\text{o}}$, a reasonable assumption for a
short-range potential.  Equating the resulting association rate,
Eq.~\eqref{eq:theta_pot}, with the rate for free diffusion in presence
of an absorbing region at $\theta = \theta_0$,
Eq.~\eqref{eq:theta_free}, we obtain $\theta_0 \approx 33^{\text{o}}$
for the width of the absorbing region, corresponding to an energy drop
of $\Delta E \equiv U(\pi)-U(\theta_0) \approx 0.4$~kcal/mol.}
\end{figure} 

For a range of parameter values we again find that an energy drop of
$\Delta E \equiv U(\pi)-U(\theta_0)$ of the order of $kT$ suffices to
enter the capture zone (Fig.~\ref{fig:rotdiff}). We find that
$\theta_0$ is insensitive to the choice of $\epsilon$ but increases as
expected with increasing $\sigma$: the range of the absorbing region
reflects the range of the potential.

\paragraph{Discussion of the toy model results.}  The toy models have
demonstrated that translational and rotational diffusional association
in a short-range potential (as it occurs in protein--protein complex
formation when the two proteins are close to each other) can be
modeled as free diffusion in the presence of absorbing regions of
suitably chosen size. The size of the absorbing regions is relatively
insensitive to the precise shape (that is, the functional form) and
magnitude of the chosen potential function; only the range of the
potential must be chosen roughly right in estimating the angular
constraints. The energy drop itself required to enter the capture zone
(binding funnel) is found to be robust towards changes in the shape,
depth \emph{and} range of the potential, and can therefore be regarded
as an essentially universal quantity that is largely independent of
the particular form of the interaction potential used in the mapping
problem.

Since our toy models employ an only one-dimensional reaction
condition, \ie a constraint on a single degree of freedom, the
question arises to what extent the relative influence of the potential
on the reaction rate would change in the case of higher-dimensional
reaction conditions (as used in our subsequent treatment of
protein--protein interactions where we impose constraints on $r$,
$\theta_{A,B}$, $\delta\phi$ and $\delta\chi$). The results obtained
by Zhou (1997) suggest that the influence of the interaction
potential on the association rate constant is more significant for the
case of two diffusing spheres bearing a circular reactive patch on
each surface (\ie where a two-dimensional reaction condition is used
for both spheres) than for the situation where one of the spheres is
taken to be uniformly reactive (\ie where a two-dimensional reaction
condition is imposed on one sphere, but an only one-dimensional
reaction condition is employed for the second sphere). Generalizing
these findings, we may anticipate that an attractive interaction
potential will affect reaction rates to a larger extent when the
number of constrained variables in the reaction condition is
increased. Since our toy models have shown that it suffices to enter
the potential well by a relatively small amount to be ``captured'', we
can conclude that for the case of a higher-dimensional reaction
condition as considered in the following, an even smaller energy drop
will be sufficient to enter the capture zone. From the point of view
of transition state theory, our approach corresponds to identifying
the transition region and then computing the flux into this region.

\subsection{Mapping the protein--protein interaction funnel from
  the structure of a protein--protein complex} 

Now we shall apply the idea outlined above to an estimate of the
angular constraints $\theta_A^0$, $\theta_B^0$, $\delta\phi_0$, and
$\delta\chi_0$, needed for the application of our expression for the
rate constant, Eq.~\eqref{kon}. For this purpose, we have directly
taken the three-dimensional structures of the considered complexes
from the Protein Data Bank (PDB).

First, the side chains of the native complexed structure were repacked
by minimizing a full-atom energy function $\mathcal{E}$ dominated by
Lennard--Jones interactions, an orientation-dependent hydrogen bond
potential, and an implicit solvation model (Gray et~al.\ 2003). As
with all current potential functions for macromolecules, there are
likely to be considerable inaccuracies in this model, but it should be
emphasized that the angular constraints and rates computed here are
relatively insensitive to the details of the interactions---the toy
examples clearly demonstrate that once the binding funnel has been
entered (which has been found to require only a small drop down in
energy), the detailed form of the interaction potential has only
little influence.

Second, a set of 1,000 alternative structures was generated from the
native complex by performing random small perturbative movements
around the native conformation, and the interaction energy of these
structures was evaluated using the same energy function $\mathcal{E}$
as employed in the repacking procedure.  The energy landscapes defined by these
alternative structures exhibit clear funnels around the native minimum.

The toy model calculations show that diffusion in such landscapes can
be modeled as free diffusion with an effective ``capture'' region
several $kT$ into the energy funnels. To define the capture energy
cutoff $\mathcal{E}_{\text{c}}$ below which the partners are committed
to bind, we compute the average $\mathcal{E}_{\text{av}}$ of the
energies of the five lowest lying structures greater than 10 \AA\ root
mean square deviation (rmsd) from the native complex (and hence
outside of the native energy funnel).  Because the energy cutoff
cannot be determined exactly, we obtain two different estimates of the
association rate setting $\mathcal{E}_{\text{c}}$ to either
$\mathcal{E}_{\text{av}}$ or $\mathcal{E}_{\text{av}}-5kT$.  We
selected the 10 structures with the largest values of
$\theta_A^0+\theta_B^0$ in the set of structures with $\mathcal{E} <
\mathcal{E}_{\text{c}}$ and took the averages of their values of
$\theta_A^0$, $\theta_B^0$, $\delta\phi_0$, and $\delta\chi_0$ to
obtain estimates for the angular tolerances used in computing the
association rates.

\begin{figure}
{\bf A}\includegraphics[angle=-90,scale=0.329]{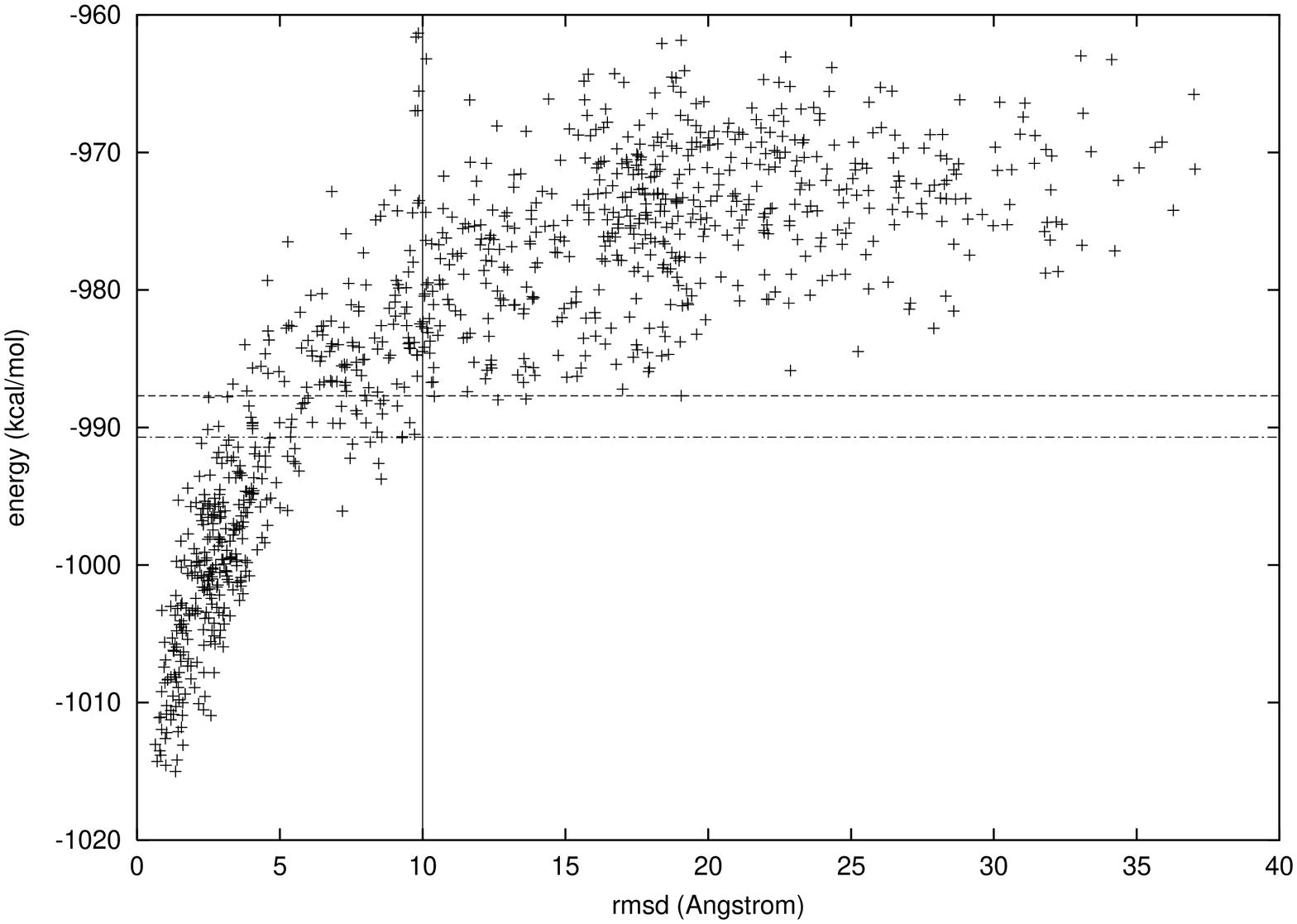}\\[.15cm]
{\bf B}\includegraphics[angle=-90,scale=0.329]{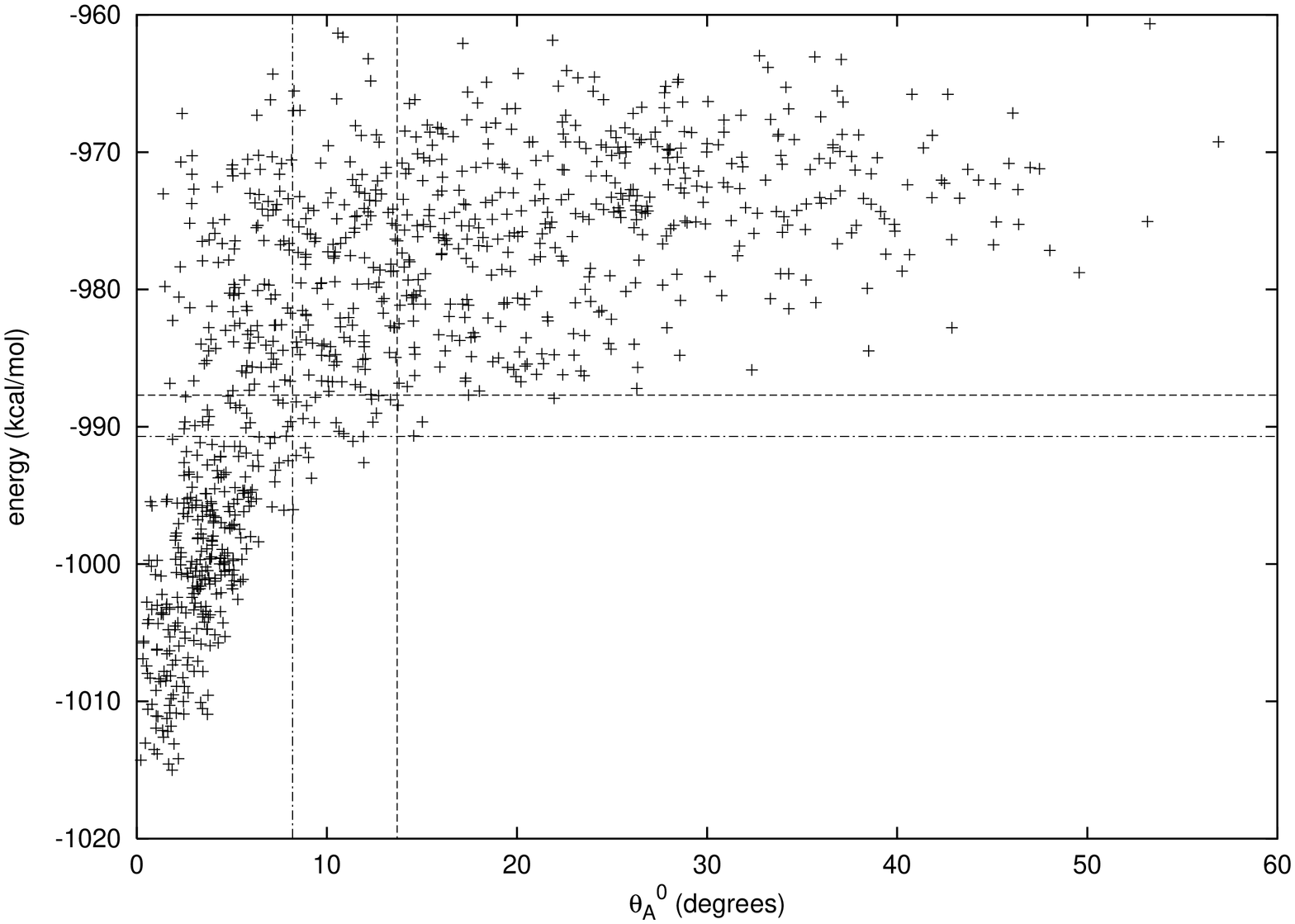}\\[.15cm]
{\bf C}\includegraphics[angle=-90,scale=0.329]{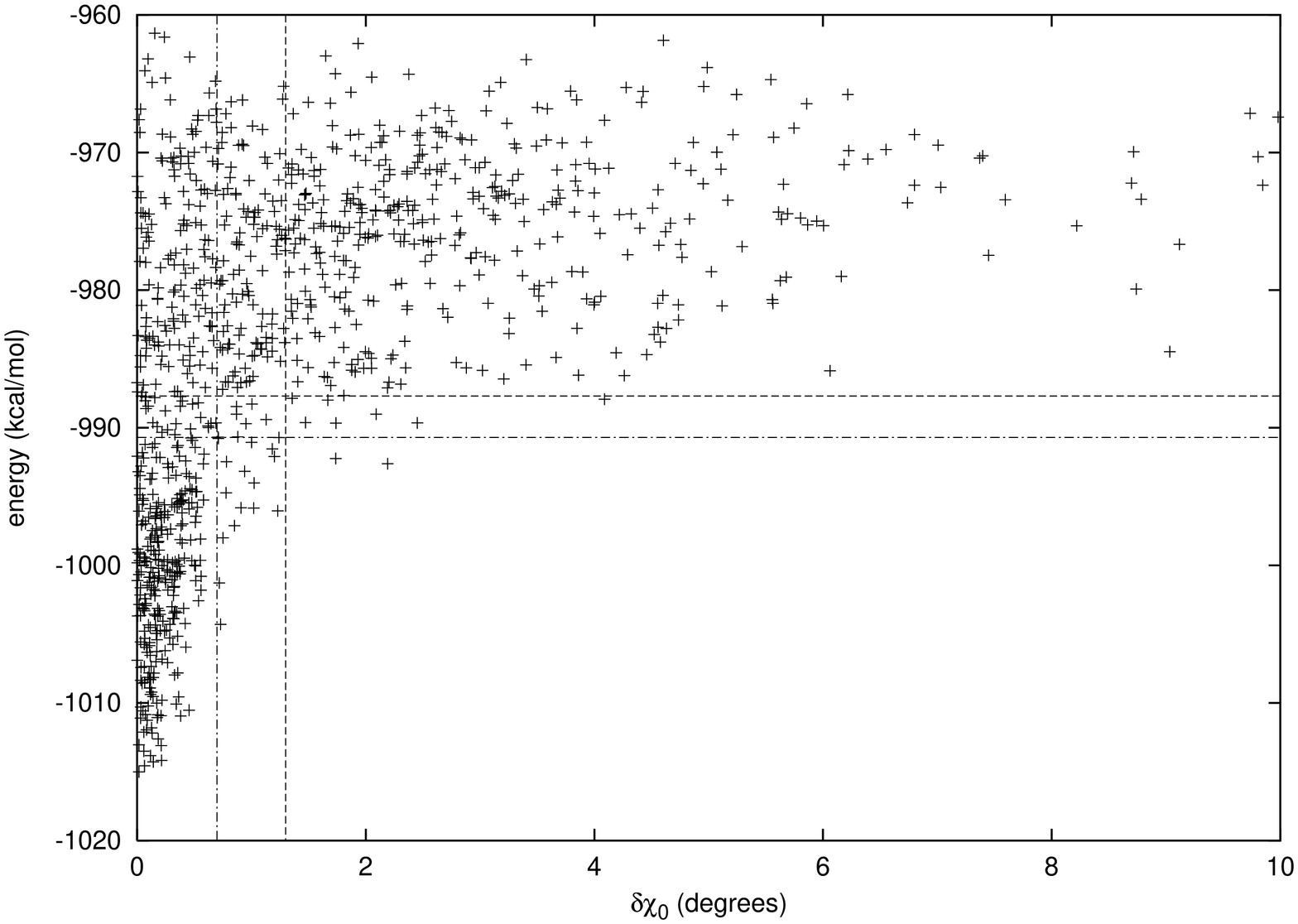}
\caption{\label{fig:1fin}
  Free energy funnels around the native structure. The energy
  $\mathcal{E}$ and rmsd (A), the energy $\mathcal{E}$ and the angular
  deviations $\theta_A^0$ (B) and $\delta\chi_0$ (C) are shown for a
  set of randomly perturbed structures of the protein--protein complex
  1FIN. States of lower energy are seen to be associated with smaller
  angles, suggesting that the angles are a reasonable measure for the
  deviation from the correctly complexed structure.  The two parallel
  lines represent the two energy cutoffs
  $\mathcal{E}_{\text{c}}=\mathcal{E}_{\text{av}}$ and
  $\mathcal{E}_{\text{c}}=\mathcal{E}_{\text{av}}-5kT$, where
  $\mathcal{E}_{\text{av}}$ is the average energy of the five lowest
  energy complexes with a rmsd above 10~\AA.  The vertical lines in
  the plots indicate the resulting angular constraints $\theta_A^0$
  and $\delta\chi_0$ corresponding to
  $\mathcal{E}_{\text{c}}=\mathcal{E}_{\text{av}}$ (dashed line) and
  $\mathcal{E}_{\text{c}}=\mathcal{E}_{\text{av}}-5kT$ (dotted--dashed
  line).}
\end{figure}

An example is shown in Fig.~\ref{fig:1fin}. We see that the
funnel-like dependence of the energy on the rmsd and on the angular
deviations is akin to the shape of the attractive potentials used in
the toy models for translational and rotational diffusion discussed
above. Furthermore, the location of the angular constraints resembles
the position of the absorbing regions of the toy models.  These
similarities support our approach of deriving the angular constraints
$\theta_A^0$, $\theta_B^0$, $\delta\phi_0$, and $\delta\chi_0$ from
the interaction energy of perturbed protein complex structures in
general, and from our method of choosing suitable energy cutoffs in
particular.

\subsection{Computation of diffusion-limited association rates
  from structures of protein--protein complexes}

\begingroup
\squeezetable
\begin{table*}
\begin{center}
\begin{tabular}{ccccccccccc}\hline\\
&&&&&&&&& \multicolumn{2}{c}{\underline{\phantom{2.5cm}$\kon$ (M$^{-1}$
    s$^{-1}$)\phantom{2.5cm}}} \\[.3cm] 

PDB & protein 1 & $R_A$ [\AA] & protein 2 & $R_B$ [\AA] & $\theta_A^0$ &
    $\theta_B^0$ &
$\delta\phi_0$ & $\delta\chi_0$ & calculated & geometric \\ \hline\\[0.2cm] 

1AVW & Porcine Pancreatic & 20.2 & Soybean Trypsin &
18.7 & 8.8 & 6.1 & 4.0 & 2.9 & $8.5 \times 10^4$ & $4.4 \times 10^1$ \\ 
& Trypsin & & Inhibitor & & 4.4 & 2.0 & 1.8 & 2.0 & $2.2 \times 10^4$ & $3.7 \times 10^{-1}$ \\[0.2cm]

1BTH & Human $\alpha$-Thrombin & 22.6 & Haemadin & 13.8 & 9.5 & 4.7
& 4.5 & 9.1 & $1.8 \times 10^5$ & $1.2 \times 10^2$ \\
&&&&& 5.6 & 2.1 & 1.3 & 8.4 & $6.0 \times 10^4$ & $2.1$ \\[0.2cm]

1DFJ & Ribonuclease A & 31.8 & Ribonuclease Inhibitor & 18.1 &
    9.3 & 8.6 & 40.9 & 19.6 & $9.8 \times 10^5$ & $7.4 \times 10^3$ \\  
     &                &       &                        &       & 5.8 &
    6.5 & 29.6 & 10.9 & $3.3 \times 10^5$ & $6.6 \times 10^2$ \\[0.2cm] 

1EFU & Ef-Tu & 30.0 & Ef-Ts & 30.7 & 11.4 & 3.3 & 2.3 & 8.2 & $1.6
\times 10^5$ & $3.5 \times 10^1$ \\
&&&&& 8.6 & 1.7 & 0.8 & 5.1 & $6.9 \times  10^4$ & $1.2$ \\[0.2cm]

1FIN & Cyclin-Dependent & 24.9 & Cyclin A & 22.9 & 13.7 & 5.8 & 
6.1 & 1.3 & $1.9 \times 10^5$ & $6.6 \times 10^1$ \\
& Kinase 2 &&&& 8.2 & 4.2 & 3.4 & 0.7 & $6.0 \times 10^4$ & $3.9$ \\[0.2cm]

1FSS & Acetylcholinesterase & 28.2 & Fasciculin-II & 14.1 & 8.8 & 10.6
& 4.2 & 9.0 & $3.0 \times 10^5$ & $4.9 \times 10^2$ \\
&&&&& 4.2 & 6.5 & 3.1 & 6.2 & $8.0 \times 10^4$ & $2.1 \times 10^1$ \\[0.2cm]

1GOT & Gt--$_\alpha$/Gi--$_\alpha$ Chimera & 26.3 & Gt--$^{\beta,\gamma}$
& 27.9 & 7.5 & 4.1 & 2.5 & 4.1 & $5.8 \times 10^4$ & $1.3 \times 10^1$\\
&&&&& 1.4 & 0.6 & 0.4 & 0.9 & $1.2 \times 10^4$ & $3.4 \times 10^{-4}$ \\[0.2cm]

1MAH & Acetylcholinesterase & 28.3 & Fasciculin-2 & 14.2 & 7.0 & 5.1
& 5.2 & 0.8 & $8.6 \times 10^4$ & $7.9$  \\ 
     &                      &       &              &       & 4.0 & 3.6
     & 1.2 & 0.2 & $3.0 \times 10^4$ & $7.4 \times 10^{-2}$ \\[0.2cm] 

1SPB & Subtilisin Bpn' & 20.6 & Subtilisin Bpn' & 15.1 &
16.9 & 17.3 & 10.9 & 21.1 & $2.0 \times 10^6$ & $2.6 \times 10^4$ \\
 & Prosegment &&&& 12.8 & 11.1 & 7.2 & 13.9 & $5.6 \times 10^5$ & $2.7 \times 10^3$ \\[0.2cm]

1STF & Papain & 20.4 & Papain Inhibitor &
16.4 & 8.4 & 4.4 & 3.9 & 1.2 & $6.9 \times 10^4$ & $8.6$ \\
&&& Stefin B && 7.7 & 4.0 & 3.0 & 1.0 & $5.4 \times 10^4$ & $3.8$ \\[0.2cm] 

1TGS & Trypsinogen & 20.2 & PSTI & 13.5 & 9.1 & 4.9 & 9.4 & 4.2 & $1.6
\times 10^5$ & $1.1 \times 10^2$  \\ 
     &             &       &      &       & 4.7 & 3.5 & 4.1 & 2.3 &
     $3.6 \times 10^4$ & 3.5 \\[0.2cm]

2SIC & Subtilisin BPN' & 20.9 & Streptomyces Subtilisin &
16.6 & 12.1 & 18.9 & 4.2 & 6.6 & $5.7 \times 10^5$ & $1.9 \times 10^3$ \\
&&& Inhibitor && 7.0 & 13.9 & 4.4 & 5.5 & $2.3 \times 10^5$ & $3.1 \times 10^2$ \\[0.2cm]

2TEC & Thermitase & 21.0 & Eglin-C & 13.9 & 8.2 & 6.5 & 2.5 & 9.0 &$1.4 \times 10^5$ & $8.9 \times 10^1$ \\
&&&&& 4.1 & 4.1 & 2.7 & 2.5 & $3.2 \times 10^4$ & 2.6 \\[0.2cm] 

3HHR & Human Growth & 26.1 & Human Growth &
21.2 & 13.6 & 14.8 & 1.9 & 3.2 & $2.9 \times 10^5$ & $3.3 \times 10^2$ \\
& Hormone && Hormone Receptor && 10.1 & 12.7 & 2.0 & 0.7 & $1.5 \times 10^5$ &
$3.1 \times 10^1$ \\[0.2cm] 

4HTC & Hirudin & 21.9 & Thrombin & 20.0 & 9.3 & 7.4 & 3.1 & 2.8 & $9.5
\times 10^4$ & $5.5 \times 10^1$  \\ 
     &         &       &          &       & 8.5 & 4.9 & 3.9 & 4.3 &
     $8.2 \times 10^4$ & $3.9 \times 10^1$ \\[0.2cm]\hline
\end{tabular}
\end{center}
\caption{
  Association rates computed from Eq.~\eqref{kon} in the fully
  diffusion-controlled limit ($\kappa \rightarrow \infty$) for the set
  of investigated protein--protein complexes. The radii $R_A$ and
  $R_B$ of each protein in the complex were estimated based on the
  radius of gyration. The angular constraints $\theta_A^0$,
  $\theta_B^0$, $\delta\phi_0$, and $\delta\chi_0$ were determined as
  described in the Methods. The geometric rates, shown for comparison,
  are given by $\kon = 4\pi D R \times \delta\phi_0 \delta\chi_0
  (1-\cos\theta_A^0) (1-\cos\theta_B^0) / 4\pi^2$.}
\end{table*}
\endgroup

With the angular constraints determined as described in the preceding
Sec.~III, we use Eq.~\eqref{kon} in the fully diffusion-limited limit
($\kappa \rightarrow \infty$) to compute protein--protein association
rates. The effective radii $R_A$ and $R_B$ of the spheres representing
the proteins are taken to be equal to the radius of gyration, $R_g =
(1/N) \sum_i d_i$ (where $N$ is the number of atoms in the protein and
$d_i$ the distance of the $i$-th atom from the geometric center of the
protein), multiplied by a correction factor of $(5/3)^{1/2}$ to obtain
the desired result $R_g = R_s$ for the limiting case of a homogeneous
sphere of radius $R_s$.  The sum $R_A+R_B$ is used as the value for
the distance $R$ between the centers of the two proteins at which
reaction is assumed to occur, \cf Eqs.~\eqref{rc}. The study of our
toy model for translational diffusion in Sec.~II has shown that this
serves as a good estimate for $R$ since the presence of short-range
attractive interactions increases the effective reaction radius only
slightly. The translational and rotational diffusion constants $D =
D_A^{\text{trans}}+D_B^{\text{trans}}$ and $D_{A,B}^{\text{rot}}$,
respectively, are computed from the Stokes--Einstein relations
$D_{A,B}^{\text{trans}} = k_B T / 6 \pi \eta R_{A,B}$ and
$D_{A,B}^{\text{rot}} = k_B T / 8\pi\eta R^3_{A,B}$, with $\eta = 8.9
\times 10^{-4}$~Ns/m$^2$ (water) and $T = 300$~K. 

Table~I lists the 15 investigated protein--protein interactions,
together with the estimated effective radii $R_A$ and $R_B$, the
angular orientational constraints $\theta_A^0$, $\theta_B^0$,
$\delta\phi_0$, and $\delta\chi_0$, and the association rate constants
$\kon$ determined from our theoretical expression, Eq.~\eqref{kon}.
For comparison, we also state the association rates obtained from a
purely probabilistic model (geometric rates).

First of all, it is worth noting that both the angular tolerances and
the corresponding rate constants are relatively insensitive (given the
approximations involved) to the particular choice of the energy
cutoffs $\mathcal{E}_{\text{c}}=\mathcal{E}_{\text{av}}$ and
$\mathcal{E}_{\text{c}}=\mathcal{E}_{\text{av}}-5kT$. For the protein
complexes under study, the rates computed from the two energy cutoffs
vary in average by a factor of 3, and no rates differ by more than a
factor of 5 for a given complex, thus indicating the robustness of our
method of estimating these rates.

We observe that the angular constraints vary significantly among the
investigated complexes, which suggests that our procedure of
estimating these tolerances yields indeed characteristic and
distinguishable values.

The association rate constants obtained using these angular
constraints range from $10^4$--$10^6$~M$^{-1}$~s$^{-1}$ and are
significantly higher than the corresponding geometric rates. While the
experimentally determined association rates of many protein--protein
complexes are in this range, considerably faster rates are also
observed, likely because of significant long-range interactions
neglected in our model.

\section{Discussion} 

We have presented a simple model for the association of proteins. The
molecules are modeled as diffusing spheres, no forces are assumed to
act between them, and the reaction condition is based on an estimate
of angular constraints on the mutual orientation of the molecular
interfaces based on the assumption of short-range guiding forces. This
procedure allows for an application of an explicit mathematical
expression for the association rate constant that we have previously
derived (Schlosshauer and Baker 2002). In this paper, we have used
this method to estimate association rates of a set of 15 different
protein--protein complexes.

The computed rates all lie within $10^4$--$10^6$~M$^{-1}$~s$^{-1}$,
which can thus be taken as the typical diffusion-limited
protein--protein association rate in the absence of attractive
interactions, in good agreement with what is experimentally known for
such interactions.  This is several orders of magnitude higher than
the geometric rate that had previously been used by various authors.
Our result therefore shows that typical diffusion-limited association
rates of proteins where no or or only weak long-range interactions are
present can essentially be explained with a model that is solely based
on translational and rotational diffusion. Experimentally observed
significantly higher rates typically suggest the presence of
electrostatic steering forces, whereas much lower rates may indicate a
reaction that is opposed by free energy barriers and is thus not fully
diffusion-limited. 

The advantage of our method over the traditional approach of BD
simulations lies in the fact that our technique provides a more
physically transparent insight into the resulting association rates.
The differences in rates among protein complexes can directly be
traced back to the sizes and shapes of the respective reactive zones
in configurational space, which are determined by mapping out the
binding funnel in the interaction energy landscape.

The model completely neglects possible free energy barriers due to
desolvation and/or side-chain freezing during complex formation as
well as a possible slowing down of diffusion within the binding funnel
due to increased ruggedness of the landscape.  Our finding that the
association rates obtained with the simple diffusional model are in
the range of those of many protein-protein complexes
($10^5$--$10^6$~M$^{-1}$~s$^{-1}$) suggests that free energy barriers
and landscape ruggedness do not have a significant impact on the
dynamics of protein-protein association.

Our model provides a zeroth-order estimate of protein--protein
association rates in the absence of long-range interactions.  This
contrasts with most previous work, which has sought to account for
changes in association rates accompanying sequence changes, rather
than the absolute association rate.  By incorporating long-range
electrostatic interactions into our diffusional model, it should be
possible to develop a complete theory of association kinetics that can
account for both the sequence dependence and the absolute magnitude of
protein--protein association rates.
 
\section*{Online ressources}

For the interested reader who would like to compute protein--protein
association rates from our model, we have created a web application
that allows the user to submit the set of reaction conditions
parameters, Eqs.~\eqref{rc}, and then returns the corresponding
binding rate by evaluating Eq.~\eqref{kon}.  The webserver can be
accessed via \texttt{http://tools.bakerlab.org/}$\sim$\texttt{pprate}.

\section*{Acknowledgments} 

We would like to thank Chu Wang for producing the sets of alternative
docked structures, and Jeffrey Gray for providing us with the energy
function used in the evaluation of these structures.  We are indebted
to H.~X.~Zhou for valuable discussions. This work was supported by a
grant from the National Institute of Health.

\section*{References}

\newenvironment{nolabellist}{\begin{list}{}
    {\setlength{\topsep}{1pt}\setlength{\leftmargin}{0.5cm}
      \setlength{\listparindent}{-0.5cm} 
      \setlength{\itemsep}{1pt}}\rm}{\end{list}}

{\small \begin{nolabellist}

\item Camacho, C.J., Kimura, S.R., DeLisi, C., and Vajda S.\ 2000.
  Kinetics of desolvation-mediated protein--protein binding.
  \textit{Biophys.\ J.\ }\textbf{78:} 1094--1105.

\item Debye, P.\ 1942. Reaction rate in ionic solutions.
  \textit{Trans.\ Electrochem.\ Soc.\ }\textbf{82:} 265--272.

\item Ermak, D.L., and McCammon, J.A.\ 1978. Brownian dynamics with
  hydrodynamic interactions. \textit{J.\ Chem.\ Phys.\ }\textbf{69:}
  1352--1360.

\item Gabdoulline, R.R., and Wade, R.C.\ 1997. Simulation of the
  diffusional association of barnase and barstar.  \textit{Biophys.\ 
    J.\ }\textbf{72:} 1917--1929.

\item Gabdoulline, R.R, and Wade, R.C.\ 2001. Protein--protein
  association: Investigation of factors influencing association rates
  by Brownian dynamics simulations.  \textit{J.\ Mol.\ Biol.\ 
  }\textbf{306:} 1139--1155.
  
\item Gabdoulline, R.R, and Wade, R.C.\ 2002. Biomolecular diffusional
  association.  \textit{Curr.\ Opin.\ Struct.\ Biol.\ }\textbf{12:}
  204--213.

\item Gray, J.J., Moughon, S., Wang, C., Schueler-Furman, O., Kuhlman,
  B., Rohl, C.A., and Baker, D.\ 2003. Protein--protein docking with
  simultaneous optimization of rigid-body displacement and side-chain
  conformations. \textit{J.\ Mol.\ Biol.\ }\textbf{331:} 281--299.

\item Janin, J.\ 1997. The kinetics of protein--protein recognition.
  \textit{Proteins} \textbf{28:} 153--161.

\item Northrup, S.H., and Erickson, H.P.\ 1992. Kinetics of
  protein--protein association explained by Brownian dynamics computer
  simulation. \textit{Proc.\ Natl.\ Acad.\ Sci.\ USA }\textbf{89:}
  3338--3342.

\item Schlosshauer, M., and Baker, D.\ 2002. A general
  expression for bimolecular association rates with orientational
  constraints.  \textit{J.\ Phys.\ Chem.\ B.\ }\textbf{106:}
  12079--12083. 

\item Schreiber, G., and Fersht, A.R.\ 1996. Rapid, electrostatically
  assisted association of proteins.  \textit{Nature Struct.\ Biol.\ 
  }\textbf{3:} 427--431.

\item Selzer, T., and Schreiber, G.\ 1999. Predicting the rate
  enhancement of protein complex formation from the electrostatic
  energy of interaction. \textit{J.\ Mol.\ Biol.\ }\textbf{287:}
  409--419.

\item Shoup, D., Lipari, G., and Szabo, A.\ 1981. Diffusion-controlled
  bimolecular reaction rates.  \textit{Biophys.\ J.\ }\textbf{36:}
  697--714.

\item Smoluchowski, M.V.\ 1917.  Versuch einer
  mathematischen Theorie der Koa\-gu\-la\-tions\-kine\-tik kolloider
  L\"{o}sungen. \emph{Z.\ Phys.\ Chem.\ }\textbf{92:} 129--168.

\item Vijayakumar, M., Wong, K.-Y., Schreiber G., Fersht, A.R., Szabo,
  A., and Zhou, H.-X.\ 1998. Electrostatic enhancement of
  diffusion-controlled protein--protein association: Comparison of
  theory and experiment on barnase and barstar. \textit{J.\ Mol.\ 
    Biol.\ }\textbf{278:} 1015--1024.

\item Zhou, H.X.\ 1997. Enhancement of protein--protein association
  rate by interaction potential: Accuracy of prediction based on local
  Boltzmann factor. \textit{Biophys.\ J.\ }\textbf{73:} 2441--2445.

\end{nolabellist}}

\end{document}